# Challenges and Recommendations in Establishing National Human Diversity Genomic Projects

**Taras K. Oleksyk***
Department of Biological Sciences,
Oakland University,
Rochester, MI 48309 USA
Department of Biology,
and
Department of Biological Sciences,
Uzhhorod National University,
Uzhhorod 88000, Ukraine
oleksyk@oakland.edu
ORCID: https://orcid.org/0000-0002-8148-3918

**Walter W. Wolfsberger***
Department of Biological Sciences,
Oakland University,
Rochester, MI 48309 USA
wwolfsberger@oakland.edu
ORCID: https://orcid.org/0000-0003-0980-645X

**Karishma Chhugani**
Department of Clinical Pharmacy,
Alfred E. Mann School of Pharmacy and Pharmaceutical Sciences,
University of Southern California,
Los Angeles, CA 90089 USA
chhugani@usc.edu
ORCID: https://orcid.org/0000-0001-8764-9352

**Yu-Ning Huang**
Department of Clinical Pharmacy,
Alfred E. Mann School of Pharmacy and Pharmaceutical Sciences,
University of Southern California,
Los Angeles, CA 90089 USA
yuninghu@usc.edu
ORCID: https://orcid.org/0000-0003-1697-4267

**Valerii Pokrytiuk**
valer.pok@gmail.com
National Pirogov Memorial Medical University,
Vinnytsya, Ukraine
ORCID: https://orcid.org/0009-0006-1355-8312

**Khrystyna Shchubelka**
Department of Biological Sciences, Oakland University,
Rochester, MI 48309 USA
Department of Biology,
Uzhhorod National University,
88000 Uzhhorod, Ukraine
kshchubelka@oakland.edu
ORCID: https://orcid.org/0000-0002-7630-3257

**Alex Zelikovsky**
Department of Computer Science,
Georgia State University,
Atlanta, GA 30302 USA
Department of Biomedical Sciences
College of Medicine and Biological Sciences
University of Suceava, Suceava,
Romania, 720229, Romania
alexz@gsu.edu
ORCID: https://orcid.org/0000-0003-4424-4691

**Bogdan Pasaniuc**
Center for Computational Biomedicine,
Institute for Biomedical Informatics,
Perelman School of Medicine,
University of Pennsylvania,Philadelphia, PA
bogdan.pasaniuc@gmail.com
ORCID: https://orcid.org/0000-0002-0227-2056

**Viorel Jinga, MD, PhD**
Department of Urology, Carol Davila University of Medicine and Pharmacy,
050474, Bucharest, Romania
Academy of Romanian Scientists,
050085, Bucharest, Romania
viorel.jinga@umfcd.ro
ORCID: https://orcid.org/0000-0001-7632-5328

**Octavian Bucur, MD, PhD**
Genomics Research and Development Institute,
050474, Bucharest, Romania
octavian.bucur@genomica.gov.ro
ORCID: https://orcid.org/0000-0002-8501-588X

**Scott C. Edmunds**
GigaScience Press,
BGI Hong Kong Tech Co. Ltd.,
Hong Kong SAR
scott@gigasciencejournal.com
ORCID: https://orcid.org/0000-0001-6444-1436

**Heinner Guio**
INBIOMEDIC Research and Technological Center,
Jose de la torre Ugarte 166, Lima 15073, Perú
Heinnerguio@gmail.com
ORCID: https://orcid.org/0000-0003-0078-1188

**Zane Lombard**
Division of Human Genetics,
National Health Laboratory Service, and School of Pathology,
Faculty of Health Sciences, University of the Witwatersrand,
Johannesburg, 2000, South Africa
Zane.Lombard@wits.ac.za
ORCID: https://orcid.org/0000-0002-7997-2616

**Brenna M. Henn**
Department of Anthropology and Center for Population Biology,
University of California - Davis,
Davis, CA 95616 USA
bmhenn@ucdavis.edu
ORCID: https://orcid.org/0000-0003-4998-287X

**Andrei Lobiuc**
Department of Biomedical Sciences
College of Medicine and Biological Sciences
University of Suceava, Suceava,
Romania, 720229
andrei.lobiuc@usm.ro
ORCID: https://orcid.org/0000-0001-5854-8261

**Alexei Levitchi**
Laboratory of Genetics,
Nicolae Testemitanu State University of Medicine and Pharmacy,
Chisinau, MD2004, Moldova
alexei.levitchi@usmf.md
ORCID: https://orcid.org/0000-0003-1784-654X

**Dumitru Ciorba**
Department of Computers, Informatics and Microelectronics,
Technical University of Moldova,
Chisinau, 2045, Moldova
dumitru.ciorba@fcim.utm.md
ORCID: https://orcid.org/0000-0002-3157-5072

**Viorel Bostan**
Department of Computers, Informatics and Microelectronics, Technical University of Moldova, Chisinau, 2045, Moldova
viorel.bostan@adm.utm.md
ORCID: https://orcid.org/0000-0002-2422-3538

**Viorel Munteanu**
Department of Computers, Informatics and Microelectronics,
Technical University of Moldova,
Chisinau, 2045, Moldova
viorel.munteanu@lt.utm.md
ORCID: https://orcid.org/0000-0002-4133-5945

**Victor Gordeev**
Department of Computers, Informatics, and Microelectronics,
Technical University of Moldova,
Chisinau, 2045, Moldova
Faculty of Electrical Engineering and Computer Science,
"Stefan cel Mare" University of Suceava,
Suceava, 720229, Romania
victor.gordeev@lt.utm.md
ORCID: https://orcid.org/0009-0005-1052-2552

**Christian P. Schaaf**
Institute of Human Genetics,
Heidelberg University,
Heidelberg, Germany
christian.schaaf@med.uni-heidelberg.de
ORCID: https://orcid.org/0000-0002-2148-7490

**Hoh Boon-Peng**
Division of Applied Biomedical Sciences and Biotechnology, School of Health Sciences
IMU University, Kuala Lumpur,
Malaysia
Malaysia Genome and Vaccine Institute (MGVI)
National Institutes of Biotechnology Malaysia (NIBM)
Jalan Bangi, 43000 Kajang,
Selangor, Malaysia

BoonPengHoh@imu.edu.my
ORCID: https://orcid.org/0000-0001-9249-4965

**Andrés Moreno Estrada**
Laboratorio Nacional de Genómica para la Biodiversidad,
Centro de Investigación y de Estudios Avanzados del IPN,
Km 9.6 Libramiento Norte Carretera Irapuato-León, C.P. 36821,
Irapuato, Guanajuato, México
comunicacionlangebio@cinvestav.mx
ORCID: https://orcid.org/0000-0001-8329-8292

**Mihai Covasa**
Department of Biomedical Sciences
College of Medicine and Biological Sciences
University of Suceava, Suceava, 720229, Romania
Department of Basic Medical Sciences
College of Osteopathic Medicine of the Pacific,
Western University of Health Sciences,
Pomona, CA 91766, USA
mcovasa@usm.ro
ORCID: https://orcid.org/0000-0002-6266-4457

**Mihai Dimian**
Department of Computer, Electronics and Automation,
University of Suceava, Suceava, Romania
Integrated Center for Research, Development and Innovation
for Advanced Materials, Nanotechnologies,
Manufacturing and Control Distributed Systems (MANSiD),
University of Suceava, Suceava, Romania
dimian@usm.ro
ORCID: https://orcid.org/0000-0002-2093-8659

**Ulykbek Kairov**
Center for Life Sciences, National Laboratory Astana,
Nazarbayev University,
Astana city, 010000, The Republic of Kazakhstan
ulykbek.kairov@nu.edu.kz
ORCID: https://orcid.org/0000-0001-8511-8064

**Victoria M. Pak**
Emory University, School of Nursing, Atlanta, GA
Emory University, Rollins School of Public Health, Department of Epidemiology,
Atlanta, GA

vpak@emory.edu
ORCID: https://orcid.org/0000-0002-4081-8347

**Seow Shih Wee**
Precision Health Research, Singapore (PRECISE)
c/o Consortium for Clinical Research and Innovation, Singapore (CRIS)
23 Rochester Park, #06-01
Singapore 139234
shihwee.seow@precise.cris.sg
ORCID: https://orcid.org/0009-0009-1075-6745

**Charleston W. K. Chiang**
Center for Genetic Epidemiology,
Department of Population and Public Health Sciences,
Keck School of Medicine, University of Southern California,
Los Angeles, CA, USA
charleston.chiang@med.usc.edu
ORCID: https://orcid.org/0000-0002-0668-7865

**Emmanuel Nepolo**
Department of Human, Biological & Translational Sciences,
School of Medicine, University of Namibia (UNAM),
Windhoek, Khomas Region, Namibia
enepolo@unam.na
ORCID: https://orcid.org/0000-0003-4302-5606

**Matteo Pellegrini**
Department of Molecular, Cell and Developmental Biology,
University of California
Los Angeles, Los Angeles, CA, USA
matteop@mcdb.ucla.edu
ORCID:https://orcid.org/0000-0001-9355-9564

**Yosr Hamdi**
Laboratory of Biomedical Genomics and Oncogenetics
Institut Pasteur de Tunis,
University of Tunis El Manar,
Tunisia.
yosr.hamdi@pasteur.utm.tn
ORCID: https://orcid.org/0000-0002-2815-1834

**Malak S. Abedalthagafi**
Department of Pathology and Laboratory Medicine, Tufts Medical Center and Tufts University School of Medicine, Boston, MA 02111, USA

Department of Neurosurgery, Tufts Medical Center and Tufts University School of Medicine,
Boston, MA 02111, USA
Malak.Althgafi@tuftsmedicalcenter.org
ORCID: https://orcid.org/0000-0003-1786-3366

**Nicola Jane Mulder**
Computational Biology Division, Institute of Infectious Disease and Molecular Medicine,
University of Cape Town,
Cape Town, South Africa
nicola.mulder@uct.ac.za
ORCID: https://orcid.org/0000-0003-4905-0941

**Jazlyn Mooney**
Department of Quantitative and Computational Biology,
University of Southern California, Los Angeles, CA, USA
jazlynmo@usc.edu
ORCID: https://orcid.org/0000-0002-2369-0855

**Javier E. Sanchez-Galan**
Facultad de Ingenieria de Sistemas Computacionales,
Universidad Tecnológica de Panamá (UTP),
El Dorado 0819-07289, Panama
javier.sanchezgalan@utp.ac.pa
ORCID: https://orcid.org/0000-0001-8806-7901

**Sandro José de Souza**
BioME, IMD
Universidade Federal do Rio Grande do Norte, Natal,
Rio Grande do Norte, 59078-970, Brazil
DNA GTx Bioinformatics
Natal, Rio Grande do Norte, 59078-970, Brazil
sandro@neuro.ufrn.br
ORCID: https://orcid.org/0000-0002-2888-6857

**Henriette Raventós**
Escuela de Biología
Centro de Investigación en Biología Celular y Molecular (CIBCM),
Universidad de Costa Rica, San José, Costa Rica
henriette.raventos@ucr.ac.cr
ORCID: https://orcid.org/0000-0001-9423-8308

**Marina Muzzio**
Instituto Multidisciplinario de Biología Celular (IMBICE),
CONICET-CIC-Universidad Nacional de La Plata,

La Plata, Buenos Aires, Argentina
Facultad de Ciencias Naturales y Museo, Universidad Nacional de La Plata, La Plata, Buenos Aires, Argentina
marinamuzzio@gmail.com
ORCID: https://orcid.org/0000-0003-2236-2494

**Gabriela Chavarria-Soley**
Escuela de Biología
Centro de Investigación en Biología Celular y Molecular (CIBCM),
Universidad de Costa Rica, San José, Costa Rica
gabriela.chavarriasoley@ucr.ac.cr
ORCID: 0000-0002-7153-4072

**Serghei Mangul**
Director | Challenges and Benchmarking
Sage Bionetworks, Seattle, WA, USA
Department of Biological and Morphofunctional Sciences, College of Medicine and Biological Sciences, Stefan cel Mare University of Suceava, 720229 Suceava, Romania
Department of Computers, Informatics, and Microelectronics, Technical University of Moldova, 2045 Chisinau, Moldova
Department of Clinical Pharmacy, Alfred E. Mann School of Pharmacy and Pharmaceutical Sciences, University of Southern California, Los Angeles, CA 90089, USA
serghei.mangul@gmail.com
ORCID: https://orcid.org/0000-0003-4770-3443

* These authors contributed equally to this work

**Abstract**

Genomic approaches have revolutionized medical research, providing valuable insights into human physiology and disease. Despite major benefits from large collections of genomes, the lack of diversity in genomic data represents a significant challenge for advancing biomedical discovery and accessible health solutions worldwide. Establishing a national genomic project is not a one-size-fits-all endeavor, as each country presents distinct challenges and opportunities. We identify challenges in the way of obtaining and publishing data from Whole Genome Sequencing (WGS) of people in various countries, discuss the progress made by some in their efforts to study their genetic diversity, and assess the most common issues. We recognize that a successful national genome database requires addressing several major issues, including the variable awareness of the recent developments in genomics among government officials, healthcare administrators, and policymakers, the absence of regulations, and ethical considerations, the challenges in securing funding, establishing legal frameworks, and building the necessary infrastructure. By assembling a diverse team of experts across 19 countries, we aim to provide a balanced approach in our recommendations to establish national projects. Our study acknowledges and addresses major intricacies and nuances specific to various settings and regions while presenting diverse opinions of scientists from both high-resource and low-resource countries contributing to a more inclusive and globally relevant framework for advancing genomic research and its applications.

## Introduction

While country borders may not represent any specific genetic boundaries, their geographic spaces contain unique genomic features, essentially representing unique pieces of a puzzle contributing to the overall picture of human genomic diversity on a global scale. The diversity they contain has been formed as a combination of forces acting during evolutionary as well as political history acting over the millennia of our past. Although differences between human populations are comparatively miniscule, even the smallest variations may be full of important information about function that can be used to help improve public health. While various strategies were historically adopted to identify said differences, in this manuscript we focused on national projects and initiatives aimed to uncover human genomic diversity of populations based on Whole Genome Sequencing (WGS), as it captures complete genome information from the individuals, revealing common as well as the rare variants that may be overlooked by the genotyping arrays ascertained by the discovery bias.

Unfortunately, the global picture of diversity remains distorted because most of the genomic data available today originates from populations of European descent[1]. At the same time, many of the low-resource regions have little to no WGS data produced, and, subsequently, limited access to full benefits from scientific discoveries in this dynamic field. Encouraging national genome projects in these regions can address knowledge gaps and improve the distribution of research benefits in these regions, reducing healthcare disparities.

The primary objective of a national human diversity population genome project is to accurately represent genomic diversity within a given country. The availability of genomic data helps to understand demographic as well as evolutionary context of human variation, offers insights into genetic variations that are invaluable for public health and biomedical research[2], and facilitates more effective treatments tailored to population genetic profiles in each country. In addition, genome projects can impact national economies by driving innovation, enhancing local infrastructure and developing local expertise in biomedical fields[3]. However, many challenges and barriers exist that prevent establishing national human diversity projects, often in the countries and regions that may need them the most.

National diversity projects create broader impacts on society by cultivating expertise and propagating national infrastructure, creating and broadening opportunities for future applications. Current national genome projects have already benefited public health and healthcare systems in several countries, from providing population-wide insights to improving individual patient care. Numerous examples illustrate how national genomic initiatives simultaneously advance population health knowledge while providing immediate clinical benefits to individuals. An early impactful example comes from Iceland's deCODE project identified actionable population-specific variants for 4% of its population, leading to updated guidelines for treatment across their healthcare system[4]. In a different geographical and genetic context, the Saudi Human Genome Program (SHGP) led to the discoveries of local genetic variants associated with hypomyelination-related neurological disorders, Leigh syndrome, autoimmune diseases resembling lupus, Meckel-Gruber syndrome, primary microcephaly, developmental epileptic encephalopathies, and muscle glycosylation defects, enabling the development of early diagnosis, targeted therapies, and preventive measures significantly reducing the healthcare burden of inherited diseases in consanguineous populations[5]. In another example, Finland's FinnGen developed the Risteys tool for population-level genetic frequencies and disease associations, leading to more accurate risk assessments and treatment planning in the country[6]. Meanwhile, the Estonian Biobank, with over 200,000 participants, provides personalized information on disease risk, prevention, and medication compatibility[7], resulting in significant findings in areas like Wilson disease management[8] and metabolic syndrome treatment in schizophrenia[9]. Singapore has also taken significant steps to mainstream precision medicine, starting with genetic testing for familial hypercholesterolemia (FH). The country is working towards integrating genomic data into clinical care by establishing Genomic Assessment Centres (GACs) to which general practitioners can refer suspected FH cases for follow-up management. These efforts exemplify how national genomic projects can drive innovation in healthcare, improve early diagnosis, and inform public health strategies for genetic conditions[10]. Finally, the actionable findings from the UK 100,000 Genomes Project significantly reduced further healthcare expenses and ended their long diagnostic journeys for 25% of 4,660 patients with rare diseases[11,12].

Many countries have yet to implement national WGS projects, despite successes elsewhere. To address the challenges and develop recommendations that can be implemented, we put together an international team of co-authors representing 19 diverse countries from around the world. This collaboration extended to multiple institutions, nonprofit organizations, policymakers, and stakeholders worldwide to include a wide range of perspectives on the establishment of national genomic projects. We aimed to include voices representing research communities from high- and low-resource nations, following classification based on the presence of infrastructure, expertise, and access to research hardware rather than economic indicators alone[13]. The global representation within our team enabled us to provide culturally sensitive recommendations from scientific, societal, and administrative angles.

In this manuscript, we review the challenges and roadblocks that contribute to the inequality in global genomic research, and provide recommendations for effective implementation of responsible research approaches in national diversity projects. We emphasize common themes such as strategic planning, local capacity building, and the development of ethical and legal frameworks that respect participant privacy and cultural diversity. These efforts ensure that the benefits of research are tangible and equitably distributed to promote open access to genomics globally, democratizing the benefits of genomic research[14]. Specifically, we organize the main challenges and considerations for national genome projects into seven major categories, describe and make recommendations for each (**Tabe 1**).

## Challenges and considerations for national genome projects

### Awareness

The variable awareness of genomic research benefits is a significant barrier to national genomic projects (**Table 1.** Awareness), and can lead to missed opportunities for valuable insights and advancements. While presenting a potential for future economic benefits in areas of personalized medicine, job opportunities, and biotech growth, genomic research involves long-term investments and commitments, which is challenging for political environments focused on more immediate gains.

Developing policies supporting genomic research requires a nuanced understanding of its implications for healthcare, privacy, and ethics, along with public perceptions, which vary by region and cultural context. Various studies have assessed attitudes toward research participation and the sharing of genetic research data, highlighting a wide spectrum of understanding and the need for more information and education about genetic research[15]. Worldwide willingness to donate samples for genomic research is low, particularly when commercial entities are involved[16]. A recent European survey indicated that 42% of the population lacked the understanding of genomic technologies, while 36% reported difficulty accessing reliable information: despite 63% of respondents expressing curiosity about personalized medicine and 60% supporting its clinical use, hesitancy persisted due to distrust in data governance[17]. These concerns align with findings from the "Your DNA Your Say" survey, which highlights that hesitancy varies based on age, education, and marginalized identity[18]. Notably, participants in South America and South Africa demonstrated a greater willingness to share genomic data with academic researchers rather than private companies[19,20], while German stakeholders emphasized privacy risks associated with public-private partnerships. In East Asia, public sentiment generally supports precision medicine, and religious authorities do not oppose precision medicine on theological grounds[22]. For instance, a national survey found that 64% of Singaporeans are willing to share de-identified health data for approved research without re-consent, with government agencies and public institutions being the most trusted data users[21].(**Table 1**. Awareness).

Understanding the intricacies of genomic research also varies across different parts of society, each facing unique awareness-associated challenges. Understanding and support for genomic research among the general public represents another critical challenge (**Table 1**. Awareness). Citizens who do not understand the potential benefits of biomedical research might oppose a national genome project due to fears and misconceptions, or simply remain apathetic to such initiatives.

In healthcare and academia, the primary awareness barrier is the lack of a clear vision of the collaborative nature of large-scale genomic research and the integration of its outcomes. Lack of incentives to pursue genomic discoveries limits the participation rates of researchers in international scientific conferences, advanced training programs, and faculty exchange. Even project conclusion, challenges persist in integrating their data outcomes into healthcare practices and academic training, preventing valuable information from being fully utilized. It is therefore not surprising that this scientific complexity and the long-term nature of the project benefits are the main limiting factors for government officials.

**Regulations**

Without adequate regulations, sensitive genetic information privacy is at risk (Table 1. Regulations). The absence of clear authorizations and role designations impacts projects from their initial planning stage. Researchers can be disincentivized from participation without clear compliance guidelines, legal protections for research activities, defined oversight and liability protocols. Similarly, for the participants, the lack of explicit protections for their respective rights, including informed consent, data privacy, and transparent policies over the use of their samples and derived data is discouraging.

Formulating comprehensive legal regulations requires careful planning, engagement with impacted communities, establishment of protective measures, and ongoing monitoring to mitigate potential risks. The US and countries in the EU have infrastructure that can maintain complex regulations. Health Insurance Portability and Accountability Act (HIPAA): Sets standards for the privacy of health information, including genetic data, while the Genetic Information Nondiscrimination Act (GINA) provides legal protection against genetic discrimination in employment and health insurance. Any human genome research must go through IRBs to ensure ethical considerations, including privacy, consent, and safety. In the EU GDPR provides individuals with substantial control over their personal data, including genomic information, by ensuring strict consent requirements and granting the "right to be forgotten".

In several countries, comprehensive legal frameworks have been crucial protection of individual privacy and ethical considerations in genome research. South Africa has established the Protection of Personal Information Act (POPIA), which aligns closely with GDPR principles[24]. Panama's Law 81 of 2019 mandates consent-based release for any information collected in medical institutions, although it does not specifically address molecular material[25]. Similarly, Brazil has adopted the Lei Geral de Proteção de Dados Pessoais (LGPD), providing a structured approach to data privacy[26]. Singapore has implemented a moratorium on genetic testing for insurance purposes, to safeguard citizens from potential misuse of genetic data and aligns with global standards for data privacy and ethics[10,23]. These examples illustrate the importance of developing and implementing robust legal frameworks that support genomic research while ensuring ethical and legal compliance (**Table 1.** Legal Frameworks).

While regulatory frameworks like HIPAA, GINA, and GDPR provide useful models, many countries must begin their genomic initiatives under existing legacy legislation and often outdated regulations that were not designed around genomic research. Regulatory frameworks typically lag behind technical capabilities, creating an environment where projects must proceed, helping to shape appropriate regulations. Since regulation is developed to varying degrees, the differences in regulations across countries may create further barriers to collaboration. At the same time, reliance on external collaborators for regulatory compliance can lead to a lack of national control over genomic resources.

## Ethics

Legal frameworks and regulations are essential to ensure ethical and responsible genetic research. The ethical considerations surrounding genomic research are critical to ensuring that research practices are respectful, responsible, and inclusive. Privacy, consent, data sharing attitudes, and the potential harm to certain communities because of the use of genetic information are major topics of discussion in the international community (**Table 1.** Privacy and Consent).

Genomic research carries potential risks of harm to identifiable population groups, which might lead to legal challenges[27]. Without careful planning, respect for indigenous laws, and establishment of ethical boards, these risks can become prohibitive (**Table 1**. Potential Harm to Groups). Opposition from affected groups can significantly hinder the collection of diverse genetic samples necessary for comprehensive research. Concerns about potential harm can lead to negative public perceptions, reputational damage for the involved institutions, and lead to the withdrawal of support from key partners, funders, or political figures.

Privacy and consent issues in human genome data sharing are amplified by misunderstandings about the benefits and protections of genomic research[28–30]. Unfortunately, concerns about sharing genomic data, particularly with private companies, are notable. If specific population groups perceive potential harm, they may resist participation, leading to biased or incomplete data. This can undermine the scientific validity and utility of the research. Co-authors from South and Central America and Asia stressed the importance of participant autonomy and benefit sharing. Addressing these concerns and ensuring ethical practices are crucial, as underscored by co-authors from Panama, Brazil, Costa Rica, and Singapore (**Table 1.** Data Sharing Attitudes).

The potential real and perceived harm for Indigenous populations is based on the historical disrespect of their councils and laws and past unethical practices (**Table 1.** Indigenous Populations)[31]. Progress in research with indigenous populations is only attainable when their perspectives are respected[32]. Some communities treat their ethnic and familial identity as deeply important or sacred and raise concerns about genomic information being a potential instrument of stigmatization. Notable examples include the misuse of the blood samples from the Havasupai Tribe in the United States[33] and the unauthorized use of genetic material from the Gyaumi people of Panama for Human T-leukemia type-2 virus (HTLV-2) drug development[34], and the consent practices in the 1997 "Anhui incident" in China[35]. These violations have resonated with the public and scientific community, which led to increased scrutiny of consent practices worldwide and an increase in negative sentiment towards genomic research across multiple Indigenous groups.

Recently, "*Your DNA Your Say*" study reviewed responses gathered from 36,268 individuals across 22 countries and demonstrated varying degrees of acceptance towards public data sharing of genomic information. Familiarity with genetics and trust in data requestors increased willingness to share data, while concerns centered on sharing with private companies[18]. These concerns stem from past experiences and ethical considerations rather than a lack of understanding. Addressing them requires coordinated efforts from scientists, governments, and the public.

Economic Considerations

In low-resource countries, economic considerations of project outcomes can be a critical factor limiting national genomic projects. In some regions, urgent healthcare challenges may be prioritized over genomic research. Most of the time, the scale of genomic initiatives demands sustainable, multi-year funding, which is challenging in regions with instability or frequent budget reallocations. In addition, geopolitical tensions and legal battles may restrict access to various technologies or services for specific countries, further hindering research and development[36,37].

The scale, typically measured by the number of participants included in the national project, directly influences resource requirements and costs. Estimation of a sufficient number of individuals to uncover the details of genome diversity in a country, especially if rare mutations are of interest, should be addressed for each scenario specifically. Countries with isolated populations, or limited migration history, like Iceland, may require fewer samples, driving their sequencing costs down. In contrast, nations with complex demographic histories and multiple ethnic groups, like India or Brazil, will require significant resources to succeed. In addition, an increase in WGS-based sample size improves the ability to detect rare, endemic mutations for the population in question.

Despite the rapidly declining price of WGS, the cost of sequencing remains a concern, limiting the sample size of national projects. Short-read sequencing, the most popular strategy for current national genomic projects, offers cost-effective high-throughput data but may miss certain structural variants and repetitive regions. Long-read sequencing provides more comprehensive coverage and better detection of complex genomic regions at a higher cost per sample. When done in bulk, short-read sequencing approaches $500[38], and long-read technologies near $1,000 per sample[39], the cost remains prohibitive if hundreds of samples are sequenced. While targeted sequencing approaches are more affordable than WGS and may appeal as suitable solutions, the reduced cost comes with substantial limitations. For example, array genotyping utilizes pre-designed probes on a microarray chip containing only several million genome sites, at 10% of the cost of WGS. Focusing only on the pre-selected genomic positions of known variation leads to strong ascertainment bias, as arrays are often designed using datasets primarily based on European populations, limiting their effectiveness in studying novel variants or population-specific diversity.[40] Another example of a targeted approach, Whole Exome Sequencing (WES) captures only the pre-selected protein-coding regions of the genome, up to 2% of the total genome sequence, at approximately one-third the cost of WGS. While significantly more valuable for clinical diagnostics and public health applications than genotyping, this approach presents limitations for diversity research, as the changes in non-coding regions of the genome are not captured[41].

Funding models

While early international genome projects like the 1KGP[42] involved massive collaborations, national projects must address their unique needs and resources. Funding strategies can be classified as governmental, private, and hybrid, each varying in proportions of domestic and foreign funds (**Table 1**. Sources of Funding). Specifically, governmental funding is fundamental for successful national genomics projects such as Genome England, but it is not always available. When successful, this results in larger-scale, multi-sectoral, collaborative initiatives, with transparent data access approaches, like the UK 100,000 Genomes Project[43], South Korea's Korean Reference Genome Project[44], or the Mexican Biobank Project[45]. Singapore's national precision medicine programme is government-funded as part of the country's Research, Innovation, and Enterprise (RIE) strategy, which aims to develop a knowledge-based economy through sustained investment in scientific research.[10,23] The government-backed Genome Tunisia Project launched in 2019 overcame challenges such as the COVID-19 crisis, funding gaps, and data-sharing regulations by integrating stakeholders, including healthcare providers, clinicians, researchers, pharmacists, bioinformaticians, industry, policymakers, and advocacy groups[46].

Because government-funded projects rely on the use of domestic resources, it is often subject to local government regulatory restrictions. Political instability and shifts of budgetary focus due to changes in administration can hinder the support of such projects. Governments might limit or delay data sharing, as in Genome of Netherlands[47], Qatar Genome Programme[48], and The Saudi Genome Project[49,50] where legal or logistical barriers slow or restrict data access. In some cases, governments could completely remove access to any data post-publication, as in Genome Russia Project[1].

Unfortunately, this mechanism is not always available. Co-authors from Moldova and Ukraine indicated that they received no government funding for their national projects. In areas with low levels of awareness, or lower-resource regions, organizers of a genome project must explore alternative solutions, including private funding. Pharmaceutical companies, biotech firms, or private grants can offer support for national genomic projects. A notable example of this approach is the deCODE genomics in Iceland, which has faced significant critique related to data privacy concerns[51]. Privately owned deCODE successfully surveyed the genomic population diversity in Iceland within the previously well-established data privacy frameworks and regulations, despite the potential issues related to their exhaustive involvement[52,53].

While privately funded projects can be quickly implemented, they don't offer the same potential as government-funded projects for integration into healthcare and growth, and may prioritize profitability over the people's needs[54,55] especially if data is eventually used to profit the funding individual or company. The hybrid model combines multiple sources of funding and incentivized service provision. *DNA do Brasil* (DNABr) initiative is a successful example of a hybrid model project, where the University of São Paulo is collaborating with diagnostic medicine company Dasa and cloud computing platform Google Cloud to sequence 15,000 WGS samples[56]. While promising potential, hybrid projects may pose complex coordination challenges, benefit sharing and ethical concerns, especially if they involve international partners. Participants of such collaborations may present varied expectations, which complicate project management, data access, and the distribution of benefits. External financial support or services, though valuable for countries in low-resource settings, could limit the local expertise and infrastructure development, and introduce risks to data privacy and potential misuse of information.

## Collaborations and Partnerships

Collaborations between local, private, and international partners can offset individuals by leveraging the combined strengths of everyone involved. International collaborations in genomic research leverage diverse expertise, resources, and data from multiple countries, enhancing the scope and impact of research projects[57]. The major challenge here is that these partnerships must navigate different regulations and data-sharing policies across countries and require transparent frameworks that promote ethical standards and facilitate navigation through the bureaucratic challenges (**Table 1.** Collaborations and Partnerships: International Collaboration)[48]. The ongoing Bahrain Genome Project exemplifies a successful international partnership, collaborating with Harvard University to advance genomic research while building local capacity[58].

Industry collaborations provide financial resources and technological capacities for genomic research projects, but are associated with risks (**Table 1.** Industry Collaboration). This approach is particularly valuable for regions with limited resources, enabling access to international expertise and infrastructure. Private companies can provide significant financial resources, enhance the accessible expertise, and provide advanced technological capabilities, as demonstrated by the deCODE genomics project in Iceland referenced in the previous section[59]. Collaborations with industry leaders such as Illumina[61], Oxford Nanopore[62], and PacBio[63] have played a key role in advancing Singapore's national precision medicine programme. These partnerships have bolstered technological capabilities in both short- and long-read sequencing[64]. However, there are concerns that industry-supported projects might prioritize profit over public benefit, potentially compromising the focus on diverse populations and ethical data use. Surveys conducted in related fields, such as genome editing, show that while US-based participants believe that the Public-Private collaboration is effective, outside of the US there is more skepticism of its value[60].

## Infrastructure and Technological Considerations

Limited infrastructure and technological expertise is a major stumbling block to the national genome projects. Geographic constraints, including distribution of population and infrastructure, pose significant logistical challenges, particularly for data collection and sample processing. Many regions lack immediate capacity for initial surveying, sample collection, storage, and DNA extraction (**Table 1.** Lack of Infrastructure). For example, high population density centers reduce the logistical burden, while surveying vast rural regions or remote population agglomerations hinders the initiatives and requires logistical planning.

Efforts to repurpose existing facilities can face resistance from healthcare and academic administrations due to academic inertia[65] or unwillingness to change the existing direction of research at various institutions (**Table 1.** Existing Facilities). Later stages face prohibitive costs of adequate laboratory facilities, sequencing hardware, and data storage. Post-sequencing analysis also requires dedicated short-term capacity for data processing[66,67]. Without careful planning, overreliance on outsourcing might significantly reduce the domestic and global utility of the project (**Table 1.** Development).

Comparative analysis of Health Technology Assessment (HTA) systems in China, India, and South Africa highlights barriers such as limited expertise, weak health data infrastructure, and fragmented healthcare systems, underscoring the need for global collaboration, especially in the Global South[68].

# Recommendations for planning and implementation of national genomic projects

The implementation of national genome projects involves multifaceted challenges and considerations. We reviewed various roadblocks and barriers, preventing many countries, especially those with limited resources from launching national genomic projects (**Table 1**). In addition, **Table S1** provides an overview of the national genomic projects discussed in this manuscript, including those in which our co-authors have participated. The table outlines project names, their status, encountered challenges, and adopted solutions, with web links to relevant literature for additional context and further reading. Addressing challenges in establishing national genome projects requires a comprehensive strategy that includes recommendations in increasing awareness, strategic planning, establishing regulation, fostering collaborations, securing sustainable funding, building infrastructure, and providing genomic data access to researchers.

## Awareness and strategic planning

The success of national genome projects depends on increasing awareness among government officials, regulatory bodies, stakeholders, and citizens about the benefits of genomic research, emphasizing potential health benefits, economic growth, and scientific advancements. Engaging diverse stakeholders is crucial in ensuring a broad-based support system for these initiatives. Failure to represent minority populations in genomics research may lead to health disparities and challenges in developing effective treatments for these groups by the national project of a given country.

Education campaigns highlighting the achievements of successful projects can address the awareness limitations, while genomic literacy programs should help with the cultural nuances, especially in diverse communities, involving local leaders and community representatives. These should increase understanding that regional and national genomic diversity can offer the opportunity for the decision makers to effectively implement concepts of personalized medicine, improve corresponding infrastructure, human resources and expertise, and provide evidence for healthcare systems budgeting[69]. Moreover, these education campaigns should also clarify the benefits and limitations of genomic research to ensure that there is a realistic understanding, particularly in how genomic data will be used and interpreted.

Researchers should be involved in increasing the awareness of genomic research benefits, and need to foster the ability to effectively communicate complex genomic concepts to diverse audiences[70]. This requires specific training in science communication, media engagement, and policy dialogue, skills that are often not emphasized in traditional research-based training of genome scientists. Outreach efforts should involve open conversations on the practical and ethical challenges posed by genomic research, acknowledging the potential risks while highlighting the benefits. Active participation in public forums, contribution to policy discussions, and local engagement are time-consuming, but necessary activities, which should be recognized and supported by research institutions as core components of scientific work, not peripheral responsibilities.

Effective communication, supported by professionals, ensures that participants and the general public understand the importance of genomic research and the potential benefits of contributing their data. Clear communication about the implications of genomic research and its potential to benefit society, alongside the need for careful consideration of its limitations, will build trust among stakeholders and participants. Engaging experts at the organizational stage would ensure sustainability and efficiency. Multidisciplinary teams composed of specialists in genomics, bioinformatics, public health, ethics, and data management would allow planning with and execution of projects with clear goals, established realistic timelines, and measurable outcomes.

Regulations are essential to systematically address the ethical challenges national genomic projects face. Creating local ethical boards and developing comprehensive regulations that reflect local differences but align with existing laws will safeguard the rights of individuals and communities while facilitating responsible data sharing and research practices. Addressing ethical concerns builds trust and encourages participation in genomic research, while considering the protection of intellectual property promotes innovation and ensures public access to essential technologies[71,72].

The complex nature of genomic research calls for an inclusive regulatory structure specifically aimed at effective support of genomic research, implementation of legal protections, and clear role designations for participants and facilitators, such as bioinformaticians, lab coordinators, and project managers conducting day-to-day research operations. Common standards in regulation should be developed to ensure protection, respect local contexts, and encourage inclusive participation while incorporating changes as technology advances.

Regulatory foundations must address challenges related to management, data privacy, consent, and sharing of genomic data, which are crucial for maintaining effective support for the conduct of research, public trust, and ethical integrity in genomic projects[73]. Facilitators and researchers should be provided with well-defined regulations that address the topics of rights, protections, and responsibilities to encourage effective research, providing protections of intellectual property and fair recognition of contributions. Responsibilities should offer clear guidelines to ensure the integrity of the research process and protect those involved from legal and ethical pitfalls. Government regulations can provide additional protections related to participants' privacy and improve the integration of the results.

While comprehensive legal frameworks provide the foundation for genomic research, national genome projects frequently must navigate incomplete regulations by implementing essential ethical and operational guidelines that can adapt as the project develops. This strategy provides critical participant protections and allows projects to move forward, maintaining important protections for participants. Project leaders should work closely with legal experts and ethics committees to interpret existing regulations in the context of genomic research, while simultaneously advocating for necessary regulatory updates.

Several existing projects offer insight into the adaptation of genomics research to incomplete existing regulations. The lessons learned in the Saudi Genome Project's[50] initial implementation identified critical shortcomings, such as incomplete metadata collection, unclear governance structures, and issues with integration of research outcomes into healthcare. Saudi Genome 2.0 initiative iterated on these issues, through collaboration with government officials, health administrators, and policymakers, improving the quality of the data produced and its integration into existing medical practices, especially in chronic disease management. The current 2.0 project strategy focuses on localizing vaccine production, exporting, and leading innovation in biotechnology, with genomics and precision medicine[74]. Meanwhile, Romania has recognized the challenges of regionally fragmented genomics research and designated genomics as strategically important in 2020. This led to the establishment of the Genomics Research and Development Institute (GRDI; via Government Decision no. 693/2021) in 2021, which is under the authority of the Carol Davila University of Medicine and Pharmacy (CDUMP) to unify national efforts. Building on regional projects[75] and aligning with the 1+ Million Genomes Declaration[76] in 2023, Romania actively participates in key European genomics projects, such as Genome of Europe[77] and Genomic Data Infrastructure[78], through the GRDI. The launch of the national ROGEN project[79] in December 2024, led by CDUMP represents a significant step towards a unified Romanian National Network of Genomic Medicine to advance healthcare nationwide, with the development of a National Strategy on Genomic Medicine underway.

New projects should explore existing regulatory frameworks, such as Latin American Critical Ethics or the European Responsible Research and Innovation (RRI) created by the EU, adopted by various funding programs (like Horizon 2020 and Horizon Europe). Similar standards can and should be implemented in local policies at the planning stages of national genome projects[80].

Additional considerations should be extended for Indigenous populations and vulnerable minorities. There are several examples of regulations that successfully addressed these issues. Native BioData Consortium (NBDC) biorepository in the United States focused on genetic and health research solutions for the benefit of all Native communities and provide safeguards for responsible sample and data storage and reuse[81]. The consortium is involved in the growth of STEM capacity in tribal communities and in fostering trust in research[81]. The CARE Principles for Indigenous Data Governance developed by Research Data Alliance[82] emphasize collective benefits, responsibility, and ethical considerations associated with these projects, aligning with the United Nations Declaration on the Rights of Indigenous Peoples[83], and have been incorporated into projects like Standardized Data on Initiatives (STARDIT)[84]. Success stories like the the Peruvian Genome Project, which obtained authorizations from the national and regional ethics committees in addition to permits from the leaders of the native communities ("Apus"), and included 17 native and 13 admixed communities (and led to findings important for regional populations like those in Ecuador and Bolivia[85,86]), demonstrate effective implementation of these principles.

## Economic Considerations

To address economic considerations, national genomic projects have to be designed with realistic funding models, goals, and cost-optimized solutions, according to the current research and public health priorities within a country. We suggest consulting with regional leaders or neighbours with established genomic expertise for special opportunities and access to genomic sequencing technologies.

Unfortunately, we still do not have robust methods to suggest a number of samples sufficient for a national genomic project, with existing national genomic projects ranging from hundreds to hundreds of thousands samples. A national project does not need to be particularly large to provide an exhaustive catalogue of local variation. A good reference point for future projects is the Korean Genome Project (Korea1K), which demonstrated that detection of common alleles reached the point of diminishing returns at 30 samples, and completely saturated after 132 samples. At the same time, nearly all the rare mutations in this study are detected in a slightly larger set (250-300 samples)[87]. Estimates should be made on a per-project basis that take into account the local population structure, historical migration patterns, geographic features, and other factors that can affect the sampling strategies.

Whole genome sequencing analysis holds scientific value at any number of samples, and strategic decisions on a project scope are made considering budgetary constraints and available infrastructure for data storage and analysis.. While WGS study cohorts often range from dozens to hundreds of samples, like examples from Papua New Guinea[88] with 58 samples or 502 individuals from Latvia[89], large-scale initiatives, such as the UK biobank, and All of US, demonstrate the advantages of analyzing hundreds of thousands of genomes. While reduced-representation methods offer cost-saving advantages, we advocate for Whole Genome Sequencing (WGS) as the preferred approach for current projects. WGS is the only method capable of providing a comprehensive view of genomic diversity and is uniquely suited to detecting rare genetic variants across the entire genome, including non-coding regions.

## Funding models

Securing adequate funding requires exploring and often combining international, governmental, and private funding approaches with clear frameworks for benefit-sharing and open-access models. Early detection and targeted treatments have already demonstrated potential to reduce long-term healthcare costs[90,91]. Government funding of projects provides significant benefits, but it is often not available for most countries from underrepresented regions. With governmental involvement, initiatives integrate into national strategies advancing genomic science and healthcare, while developing local expertise and infrastructure. Convincing governments and investors that genomic programs lead to imminent improvements in public health and reduce healthcare costs could be the key to justify national genomic initiatives[92].

Projects with private funding offer higher flexibility and reduced bureaucratic delays, leverage existing private resources or expertise, but require the adoption of a robust regulatory system ensuring fair and equitable outcomes for genomic research. Industry collaborations offer access to cutting-edge infrastructure and expertise but require careful balancing of data access, ownership, and benefits distribution. Such partnerships should operate within transparent regulatory frameworks, supported by incentives such as tax breaks and facility sharing agreements. Allocating a portion of commercial gains from patents back into the community sustains momentum and ensures reciprocal benefits, fostering ownership among community members and translating scientific advancements into tangible public health improvements. Ongoing monitoring and evaluation of expenditures and progress ensures projects remain economically viable.

Hybrid funding can offset the limitations of single-source approaches and remains the most accessible in resource-limited regions. In cases when local expertise and resources are limited, international collaborations can provide knowledge and access to critical infrastructure. Uncertainties in terms of data privacy and ownership must be explicitly addressed and balanced in the early stages of planning, with collaborations set up as partnerships among the parties, and data sharing[93]. Alternative business models for revenue generation from national genome projects, such as incorporating direct-to-consumer genetic testing and partnerships with pharmaceutical companies, should be carefully balanced with ethical governance and transparency[94]. Collaborative efforts among policymakers, researchers, industry leaders, and the public are critical to addressing these challenges and building a sustainable, responsible genomics ecosystem.

## Collaborations and Partnerships

**International collaboration** and funding are essential for management of genomic research challenges and ensuring global access to its benefits. Fostering successful international partnerships requires establishing transparent frameworks that promote ethical standards and facilitate navigation through the bureaucratic challenges. Industry collaborations must balance access to genetic resources and benefits with transparency in partnerships. Ensuring public or governmental oversight, adopting an open-access model where data and research findings are available to the public without restriction, and maintaining transparency in such collaborations can help alleviate data privacy concerns and build trust among stakeholders  Clear agreements on data sharing, privacy, and benefit distribution can mitigate potential conflicts and enhance these efforts.

Researchers can try to extend beyond traditional grant systems to include diverse sources such as public-private partnerships and international agencies to ensure a stable financial foundation for a project.  In the absence of domestic funding, the national project in Ukraine was negotiated as a collaboration of a local research group with leading genomic groups in the US and China (NIH and BGI)[57], where international partners provided whole genome genotyping and sequencing respectively, with bioinformatic analysis carried out by the local team. Potential concerns about data privacy issues were alleviated by the project adopting a transparent, full public access model for data sharing[95]. Addressing the power disparity between the wealthy and the low-income countries requires strategic incentives beyond simple calls for diversity or equity. Financial incentives alone are often insufficient, but aligning genome research with global goals such as economic growth, scientific advancement, or public health security can create shared value for all. For instance, sequencing the underrepresented populations can yield more valuable insights than sequencing additional samples from well-studied populations, as well as alleviate current "pioneer advantage biases"[1,96]. Local genetic variations improve our understanding of rare diseases, drug responses, and resistance mechanisms, globally advancing precision medicine efforts. Worldwide efforts and threats such as the recent SARS-CoV-2 crisis demonstrate the importance of uniform understanding of the genetic variation across diverse populations and the necessity of preparedness in terms of infrastructure and expertise across all nations[97].

International organizations and high-resource countries should continue to support genomic projects through grants, incentives, loans, and access to expertise. These frameworks should be flexible enough to accommodate unique needs ensuring benefits for their participants. Global Alliance for Genomics initiative has promoted and sustained human genome projects in low-resource regions, without providing direct funding[98]. The Human Heredity and Health in Africa (H3Africa) initiative was supported by the Wellcome Trust and NIH. The Harnessing Data Science for Health Discovery and Innovation in Africa, was supported by the NIH[99], to promote data science and accelerate genomics research and deploy genomic resources to solve public health challenges in the continent[79]. The combination of mentioned support enhances the ability of scientists to built new collaborations, develop genomics research infrastructure and apply genomic approaches to to unravel the genomic predisposition of certain communicable and noncommunicable diseases[78].

Forming international consortia is crucial not only for facilitating knowledge exchange and diverse population representation in genomic databases, reducing healthcare disparities. Collaborations with Asian countries, particularly in the context of their vast and diverse populations, are equally important to ensure that global genetic research is comprehensive and inclusive. These collaborations can address critical gaps in knowledge and provide invaluable insights into the genetic basis of diseases that are prevalent in Asia. International standards for genomic research, including data collection, processing, and sharing, should promote inclusivity and equity. The All of US initiative, which produced WGS data for more than 300,000 samples[100], demonstrates the possible challenges, facing concerns over the correct application of analytic techniques and the resulting harmful social implications[101]. This underscores the need for inclusion of a diverse set of experts that consider not only the facilitation of the project but its potential impact upon analysis and release. Regional collaborations offer unique advantages for national genomic research, as neighboring countries often share similar challenges, history, and environmental contexts. The presence of even one successful story within a broader region can propagate further collaborative research within bordering countries. This makes the experience and the strategies adopted by a regional success a better template for planning a future national genomic project in the proximity. Broader support of regional initiatives leaders can be a more effective solution than the development of international collaborations[102,103].

There are examples from genomics as well as other disciplines to follow[104,105]. Advocating for a cultural shift toward open data management practices across all scientific disciplines enhances the scope and impact of scientific discovery, transforming local communities from mere subjects of research to active beneficiaries (**Table 1.** Data Access: Open Access and Data Privacy). Integration of FAIR principles into global genomic research is crucial for ensuring that communities contributing to such studies significantly benefit, especially in terms of healthcare improvements and local development[106].

However, critical concerns regarding equity in global genomics collaborations, particularly the flow of funding, capacity-building metrics, leadership representation, and the persistence of "helicopter science" in Global South partnerships persist. While our study does not directly assess these dynamics, we recognize their profound ethical and practical implications for fostering effective research ecosystems. Initiatives, such as the H3Africa Consortium and UNESCO's recommendations on open science emphasize the importance of local infrastructure investment, leadership parity, and transparent funding frameworks to mitigate extractive practices. Further empirical studies are needed to systematically evaluate progress and challenges in these areas. We agree that addressing power imbalances and prioritizing sustainable capacity-building must remain central to ethical global genomic research.

## Infrastructure

Different stages of population genomic studies present unique challenges to infrastructure. Early-stage sampling challenges can be addressed with technological solutions, such as mobile collection units, support of both web and paper-based surveying instruments, and the implementation of advanced remote sample collection protocols, while sample processing can effectively leverage existing laboratories and healthcare infrastructure, coupled with training for the current staff (**Table 1.** Existing Facilities). Still, the cost of facilities and hardware for DNA sequencing, data storage, and analysis remains prohibitive for many countries. Outsourcing to collaborative partners or private sequencing centers is an option to mitigate these costs, though it requires careful consideration of data security, property rights, and logistical coordination (**Table 1.** Outside Resources).

For data analysis, adaptation of existing local computational facilities remains the most sustainable approach. In contrast to many computational problems posed by other scientific fields, genomics requires large data storage capacity. A research-grade WGS project for 100 samples yields approximately 12 TB of raw data, requiring up to 40 TB of storage to accommodate the analysis with a single back-up. In the absence of local computational resources, advancements in modern computing hardware enable such projects to be analyzed within several months using a professional-grade workstation and network storage, with a budget of under $10,000. Lastly, internationally accessible cloud-based services or platforms like Terra, Amazon AWS HealthOmics, or Google Cloud Platform can be utilized. They offer affordable and rapid services for singular data analysis tasks, with single samples read-to-variant analysis approaching high single-digit values depending on the platform[107,108]. Unfortunately, the sustained cloud service usage is associated with extended costs due to the large data volumes associated with genomic projects at a population scale, with data storage and transfer expenses making them the least desirable option for most national genome research projects that aim to share their data widely[109]. Popular international genomic data storage platforms, such as the European Genome-Phenome

Archive (EGA)[110], NCBI Sequence Read Archive (SRA)[111], or DDBJ Japanese Genotype-phenotype Archive[112] can be used as long-term host and provide access to the project data for collaboration for smaller projects. However, these platforms are also limited in terms of the maximum data deposited. For example, EGA allows for 10 TB of data from a single submitter, with SRA limiting a single submission to 5TB. Depositing such volumes of data through the internet requires high-speed, reliable internet connections. Lastly, it is important to note that non-local solutions will result in the loss of flexibility for compliance with local regulations, data storage, and sharing.

The decisions to incorporate specific sequencing technology should be driven by cost and data yield. The international community should strive for a state where geopolitics and lobbying are not important in these decisions. Facilitators of national genomics should be provided with unbiased independent benchmarking metrics, and cross-platform omics analysis protocols to ensure the optimal course of action in terms of the selection of technological solutions for their projects[113,114].

Even after a successful implementation of a national genomic project, careful consideration has to be extended to responsible approaches to data analysis, interpretation, and presentation. While analytical methods and tools are widely available, their uncritical application without a deep understanding of methodology might undermine the overall validity and impact of findings. To choose appropriate tools for analysis, researchers should refer to systematic benchmarking studies[115], which evaluate computational methods across diverse biological contexts. Additionally, sequence analysis often involves multi-step workflows, with numerous specialized tools. complicated installation and setup, and iterative data conversion steps[116]. Standardized frameworks such as Nextflow Sarek[117], address some of these challenges by automating workflow execution and simplifying technical reproducibility. While such tools address the technical barriers, the core bioinformatics challenges, such as context-dependent parameter selection, compatibility with reference datasets, and the nuanced interpretation of results, require deep domain-specific expertise.  Sustained bioinformatics support is essential for uncovering further value from genomic data long after initial sequencing and analysis are complete. Advanced analysis and healthcare integration require further development of infrastructure and expertise. Investment in local expert training builds national scientific expertise and ensures continuity, and project sustainability supplemented by technological and academic support through international partnerships[104,118].

Genome infrastructure supports sectors related to public health and biotechnology and provides new opportunities for economic development. Therefore, national genome initiatives should prioritize developing local scientific infrastructure to support sustainable growth. Capacities inherited from these projects pave the way for the development of cost-effective treatments and enhanced disease management strategies, significantly reducing healthcare spending, and can drive new technologies and methodologies, contributing to job creation and economic growth[2].

## Genomic data access, public data sharing

Data sharing considerations vary for data on different analysis stages. Primary data sharing enables future reanalysis and broader integration, but also larger data volumes, increased cost, and computationally complex re-analysis. Secondary analysis results are more compact and can be rapidly incorporated in healthcare studies, while severely limiting reanalysis[119]. There are multiple challenges to depositing and sharing the genomic data: technological limitations, privacy and consent concerns, national security issues, and agreements with companies that have contributed data. Various efforts aim to introduce standards for data sharing in different areas of genomics, with a focus mostly on biomedical topics. As a useful resource for new national projects, we can recommend *The Global Alliance for Genomics and Health* (GA4GH) which presents toolkits, frameworks, and guidelines that can be used to provide responsible data access to genomics data[120].

The heterogeneity of genomic data-sharing platforms presents a significant challenge to genomic data sharing. While repositories like European Genome-Phenome Archive (EGA)[110], The European Nucleotide Archive (ENA)[121], and NCBI Sequence Read Archive[111] provide researchers the ability to host the data and set up access to it in a standardized fashion, many national projects use in-house solutions due to local regulations. This fragmentation hinders data integration and collaborative research.

We advocate for the application of FAIR (Findable, Accessible, Interoperable, Reusable) principles in scientific data management to enhance the accessibility and utility of genomic data globally[106]. Genomic data utility is maximized when it is available through an open-access model[122]. This means the data files are freely available to anyone without monetary cost or any other restrictions or barriers, potentially accelerating research. 1KGP international project pioneered this approach by depositing the collected genome sequences and the accompanying metadata[123,124] into a curated international database that could protect the participant interests, while serving researchers worldwide.

In contrast, most national projects today choose a restricted access model, with only the summary information publicly available, which is sometimes the result of national research policies that prohibit a complete open access model. While this approach may address privacy concerns or comply with local regulations, it limits the broader utility of the data for advancing global genomics research. To maximize impact, national projects should prioritize the integration of their data into the global knowledge base, enabling cross-border collaborations and comparative studies using data federation and emerging forms of privacy protection technology and privacy-enhancing technology. Such integration should be recognized as a key metric of success, balancing the benefits of open access with necessary safeguards for privacy and ethical use. To advance global understanding of human genome diversity, integration of data into the global knowledge base, where possible, should become the major metric of success for any national genome project.

## Conclusions

The overall success of national genomic projects relies on collaboration, interdisciplinary expertise, effective communication, and balanced regulatory, public, and private support. Investment in local genetic variation research is critical for developing precision medicine and shaping healthcare systems within each country. Our practical recommendations can be adopted by all countries to contribute to global genomics research and benefit from the knowledge gained through these efforts (**Table 1**).

To address these challenges, we call on the global community to promote inclusive genomic research that includes diverse populations and provides open access to genomic data. First, we must increase the collection and sequencing of genomes from under-studied populations. Second, we must increase public awareness of sharing and using such genomic data. However, the demand for public data from around the world has to be balanced with the rights of sovereign nations and indigenous tribal groups to control access to their genetic data. The World Health Organization (WHO) recently launched an initiative to promote fair access to genomics globally, democratizing the benefits of genomic research[14]. We advocate for the application of FAIR (Findable, Accessible, Interoperable, Reusable) principles in scientific data management to enhance the accessibility and utility of genomic data globally[106]. The open access model is the best approach for global collaborations, can provide researchers with the ability to integrate the data in future global studies, and apply updated analysis techniques[78], However, adequate regulations and considerations should protect those providing their genomic information[125].

Concerted effort from the researchers is required to ensure our goals are aligned with communities where research is occurring[31]. The establishment of more nationwide genome projects throughout the world is crucial in advancing medical research and improving public health. We call for global funds and research groups to support human genome projects in low-resource countries that are in the early stages of implementation of such projects. Investing in global genomics research is essential, and multiple separate challenges must be addressed to increase the representation of diverse populations.

# Tables

**Table 1.** Structured review of challenges, considerations, and recommendations for effective national genome projects

| Issue | Sub-category | Description | Recommendation(s) |
|---|---|---|---|
| **Awareness** | Government officials and policymakers | Variable understanding of genomic research benefits among government organizations | Educate and engage decision-makers; promote the potential benefits of genomic research. |
| | Stakeholders (facilitators) | Need to increase awareness among healthcare, academia, and industry. Lack of effective science communication. | Multi-sectoral cooperation promotes an understanding of benefits. Incentivization of science communication activities as core. |
| | Public | Low to neutral levels of public understanding and support for genomic research. | Public education campaigns; transparent communication of benefits and safeguards. |
| **Regulations** | Decision-making bodies | Absence of necessary legal frameworks to protect privacy and prevent discrimination. | Establish robust legal frameworks; create ethical boards in each country. |
| | Legal frameworks | Need for legal frameworks to support genome research. | Develop comprehensive legal regulations; consider existing laws relevant to genomic data sharing. Start with essential protections, incorporating improvements as the project evolves.. |
| **Ethics** | Privacy and consent | Concerns over privacy, consent, and use of genetic information. | Develop safeguards; consult with affected communities; and ensure transparency. |
| | Potential harm to groups | Risk of harm to identifiable population groups; legal challenges. | Careful planning; respect indigenous laws; establish ethical boards. |
| | Data sharing attitudes | Attitudes towards sharing human genomic data. | Address concerns about sharing for private profit; ensure ethical practices. |
| | Indigenous populations | Specific considerations related to indigenous populations. | Respect indigenous councils and laws; prevent a repetition of past unethical practices. |

| Issue | Sub-category | Description | Recommendation(s) |
|---|---|---|---|
| **Economic Considerations** | Scale and cost | Addressing the scale and the technologies used in a project. | Using smaller scale projects to develop expertise and infrastructure, focusing on regionally available WGS approach. |
| **Funding** | Sources of funding | Challenges in securing governmental, private, and hybrid funding approaches. | Develop clear frameworks, benefit sharing, and open-access models; enhance local capacities; promote international cooperation; |
| **Collaborations & Partnerships** | International collaboration | Challenges and opportunities in international collaboration. | Foster regional and international partnerships; navigate bureaucratic obstacles. develop equitable standards for compatibility |
| | Industry collaboration | Collaboration with the pharmaceutical industry in funding and providing expertise. | Balance access to genetic resources and benefits; ensure transparency in partnerships. balance access to genetic resources and benefits; consider open-access models. |
| **Infrastructure and Technological Considerations** | Lack of local infrastructure | Economic and geographical constraints in many countries. | Emphasize collaboration, shared resources, and data infrastructure. |
| | Existing facilities | Utilizing existing facilities and resources for genomic projects. | Repurpose or upgrade existing laboratories and healthcare infrastructure; retrain staff. |
| | Existing technologies | Adaptation and utilization of existing facilities and resources. | Repurpose or upgrade existing laboratories and healthcare infrastructure; retrain staff. |
| | Outside resources | Access to facilities and technologies outside the country | Seek collaborations and partners for shared infrastructure and expertise. |
| | New technologies and infrastructure | Challenges in developing new technologies and infrastructure. | Invest in new technologies; seek collaborations and funding. |

**Acknowledgements and Declarations**
We thank Dr. Sarah Tishkoff for her valuable feedback and discussion.
The author(s) received no financial support for the research, authorship, and/or publication of this article.

**Declaration of generative AI and AI-assisted technologies in the writing process**
During the preparation of this work, the author(s) used OpenAI ChatGPT, Google Gemini, and Grammarly. These tools were utilized solely to improve grammar, clarity, and streamline the writing process, particularly as English is a second language for most of the authors of this manuscript. They were not used to generate scientific content, figures, data analysis, or conclusions. Following the use of these tools, the author(s) reviewed and edited the content as needed and take full responsibility for the content of the publication.

**Author Contributions**
S.M. and T.K.O. developed the concept and design and supervised the study.
W.W., K.S., V.P., S.M., and T.K.O. developed and facilitated the study.
C.P.S., E.N., G.C., H.G., H.R., H.B., J.E.S-G., M.M., S.J.S., U.K., Y.H. participated in the study surveys.
W.W., K.S., and V.P. analyzed the data.
A.Z., A.L.(1), A.L.(2), A.M.E., B.P., B.H., C.W.K.C.P., D.C., V.B., G.C., J.L.R., K.C., M.C., S.C.E., W.W., V.G., V.M., S.M., and T.K.O. contributed to the writing of the initial draft of the manuscript.
W.W., E.N., G.C., H.R., J.M., K.C., M.S.A., M.M., M.P., N.J.M., S.J.S., S.G., U.K., V.M.P., V.G., Y.H., T.K.O., and S.M. contributed to the final draft of the manuscript with inputs from all authors.